\documentclass[prl,twocolumn,amsmath,amssymb]{revtex4}
\usepackage{amsmath}
\usepackage{graphicx}
\usepackage{bm}
\usepackage{hyperref}
\usepackage{epsfig}
\usepackage{amsfonts}
\usepackage{amsthm}

\begin{document}

\title{Signatures of Strong Correlations in One-Dimensional Ultra-Cold Atomic
Fermi Gases}

\author{Paata Kakashvili, S. G. Bhongale, Han Pu and C. J. Bolech }

\affiliation{Physics \& Astronomy Department and Rice Quantum Institute, Rice
University, Houston, TX 77005, USA}


\begin{abstract}
Recent success in manipulating ultra-cold atomic systems allows to
probe different strongly correlated regimes in one-dimension. Regimes such as
the (spin-coherent) Luttinger liquid and the spin-incoherent Luttinger liquid
can be realized by tuning the inter-atomic interaction strength and trap
parameters. We identify the noise correlations of density fluctuations as a
robust observable (uniquely suitable in the context of trapped atomic gases) to
discriminate between these two regimes. Finally, we address the prospects to
realize and probe these phenomena experimentally using optical lattices.
\end{abstract}
\maketitle

Strong correlations in low-dimensional Fermi systems give rise to
interesting many-body states such as those responsible for High-$T_{c}$
superconductivity and fractional quantum Hall effect (FQHE) in two
dimensions (2D), and Luttinger liquid~\cite{gogolin_bosonization_2004,giamarchi_quantum_2004}
in one dimension (1D). In particular, the Luttinger liquid phase has
been the paradigm of low-energy physics in 1D systems for about half
a century~\cite{tomonaga_remarks_1950,luttinger_exactly_1963}. This
phase is characterized by the absence of fermionic quasi-particles
even in the presence of a well defined Fermi surfaces with the relevant
modes of dispersion represented by bosonic spin and charge excitations
propagating at different velocities (spin-charge separation). Very
recently, a new regime, the spin-incoherent (SI) Luttinger liquid~\cite{cheianov_nonunitary_2004,fiete_colloquium_2007},
has become an active area of research. In contrast with the spin-coherent
(SC) Luttinger liquid, here the spin-incoherence results from the
induced spin-spin interactions scale $J=2\pi^{2}\hbar^{4}\left\langle \rho\right\rangle ^{3}/(3m^{2}U_{\text{1D}})$
\cite{matveev_conductance_2004} being the smallest scale in the system,
$J\ll T\ll E_{F}$, where $E_{F}$ is the Fermi energy, $T$ the temperature,
$\left\langle \rho\right\rangle $ the total density and $U_{\text{1D}}$
the interaction strength for contact interactions. In systems of ultra-cold
atoms, this regime can be reached by increasing either the interaction
strength among particles or by reducing the density. Thus, even at
extremely low temperatures, fluctuations may drive the Luttinger liquid
to lose spin coherence and only charge excitations remain as the dominant
propagating mode. While there exists a growing experimental evidence
in support for the SC Luttinger liquid~\cite{giamarchi_quantum_2004},
experimental evidence for the SI Luttinger liquid is very scarce~\cite{auslaender_spin-charge_2005_short,steinberg_localization_2006_short}.
Even though it is possible to change density in 1D condensed matter
systems by applying a gate voltage, lack of tunability in the interaction
strength limits possible experimental realizations. In contrast, in
trapped ultra-cold atomic systems the SI Luttinger liquid regime is
unavoidable since low-densities are inevitably reached near the confinement
edges.
\begin{figure}
\includegraphics[width=3.3in,keepaspectratio]{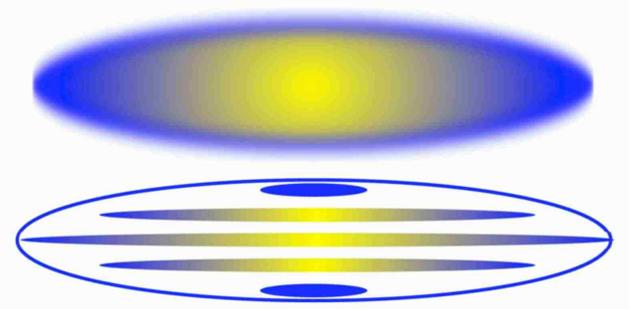} 
\caption{Schematic sketch of the 3D cigar-shaped cloud (upper)
and array of 1D tubes (lower) formed by optical trap lattice potentials.
Yellow (light) regions represent the spin-coherent Luttinger liquid regime,
while blue (dark) regions indicate its spin-incoherent counterpart.}
\label{1D_tubes} 
\end{figure}

With the recent developments in the techniques of trapping and manipulation
of ultra-cold atomic gases, the study of the strongly correlated regime
of many-body systems has acquired new momentum. Trapped atoms form
an ideal many-body system that can be configured with extreme control
and purity allowing for a thorough study of many-body properties previously
inaccessible in solid state configurations. Experimentalists are able
to achieve strongly correlated regimes either by configuring the interaction
to be strong by use of a tunable atom-atom scattering resonance, for
example a magnetic Feshbach resonance~\cite{feshbach_unified_1958},
or by introducing degeneracy in the single particle ground state by
creating a lattice trapping potential~\cite{bloch_ultracold_2005}.
The former occurs when a two-body bound molecular states is made resonant
with the open channel threshold by dialing-in an external magnetic
field. This allows for the effective two-body interaction to have
a negative (attractive) or positive (repulsive) sign with a strength
that is dependent on the distance of the bound state from the scattering
threshold. The lattice potential is induced by creating a standing
wave pattern when two laser beams interfere at an angle~\cite{bloch_ultracold_2005}.
The depth of the individual lattice well and the distance between
wells is directly related to the intensity and the wavelength of the
laser light. Such controllability made available by optical lattices
has played an indispensable role in the trapped-dilute-gas demonstration
of condensed matter phenomena such as superfluid to Mott insulator
transition~\cite{greiner_quantum_2002_short}. In other experiments, multiple
beams were used to create an optical lattice in two spatial dimensions,
for example $x$ and $y$, resulting in a configuration consisting
of an $x$-$y$ array of tubes as depicted in Fig.~\ref{1D_tubes}.
Each tube can act as an independent elongated quasi-1D trap if the
tunneling between the tubes is adjusted to be zero with the transverse
trapping frequency large compared to all scales, including interaction
energy scale. Such controlled 1D-trap arrays allowed for the successful
demonstration of the Tonks-Girardeau Gas regime~\cite{paredes_tonks-girardeau_2004_short,kinoshita_observation_2004}
and play a key role in theoretical proposals of achieving a dilute-gas
SC Luttinger liquid~\cite{recati_spin-charge_2003_combined_short,kecke_charge_2005}.
Moreover, trapped multi-species/spin atomic systems offer the added
advantage of individual addressability. This allows for measurement
of properties that are species/spin dependent. For example, it is
possible to measure the spin up-up and down-down correlation function
separately.

In this Letter, we extend the 1D Fermi gas studies to both regimes
of the Luttinger liquid in two-component quasi-1D Fermi gas as confined
in an array of tubes as shown in Fig.~\ref{1D_tubes}. Previous theoretical
studies of this regime in ultra-cold Fermi gases have been limited
to investigating the separation between charge and spin degrees of
freedom~\cite{recati_spin-charge_2003_combined_short,kecke_charge_2005}.
Here, we focus on density fluctuations and show that there is a qualitative
difference between the SC Luttinger liquid and its spin-incoherent
counterpart.

We consider a two-component ultra-cold atomic gas in a 3D magnetic
trap. To define quasi-1D tubes, an optical lattice is imposed in the
transverse direction. The combined effect of the trap potential can
be modeled by $V(\vec{x})=m\omega_{\perp}^{2}(x^{2}+y^{2})/2+m\omega_{z}^{2}z^{2}/2+V_{0}\left[\sin^{2}(2\pi x/\lambda)+\sin^{2}(2\pi y/\lambda)\right]$.
Here, $\omega_{\perp},\omega_{z}$ are the trap frequencies for the
cigar-shaped ($\omega_{z}<\omega_{\perp}$) trap, $\lambda/2$ is
the period of the optical lattice and $V_{0}$ is its depth. The interaction
between atoms is modeled by the contact pseudopotential $V_{\text{int}}(\vec{x}_{1}-\vec{x}_{2})=U_{\text{3D}}\delta(\vec{x}_{1}-\vec{x}_{2})$,
with $U_{\text{3D}}=4\pi\hbar^{2}a_{\text{3D}}/m$, where $a_{\text{3D}}$
is the 3D $s$-wave scattering length. For large depth of the optical
potential, the 3D cloud splits up into 1D tubes and each tube is then
described by the following 1D Hamiltonian:
\begin{eqnarray}
H_{\text{1D}} & = & \sum_{\sigma=\uparrow,\downarrow}\int dz\psi_{\sigma}^{\dagger}(z)\left[\frac{p_{z}^{2}}{2m}+V_{\text{trap}}(z)\right]\psi_{\sigma}(z)+\nonumber \\
 & + & U_{\text{1D}}\int dz\psi_{\uparrow}^{\dagger}(z)\psi_{\downarrow}^{\dagger}(z)\psi_{\downarrow}(z)\psi_{\uparrow}(z),
\label{eq:Hamiltonian1D}
\end{eqnarray}
where $\psi_{\sigma}^{\dagger}(\vec{x})$ {[}$\psi_{\sigma}(\vec{x})$]
is the creation {[}annihilation] operator for particles of spin $\sigma$.
$V_{\text{trap}}(z)=m\omega_{z}^{2}z^{2}/2$ is the 1D trap potential
and $U_{\text{1D}}=-2\hbar^{2}/ma_{\text{1D}}$ represents the effective
1D interaction strength defined in terms of the effective 1D scattering
length $a_{\text{1D}}=-(a_{\perp}^{2}/a_{\text{3D}})(1-Ca_{\text{3D}}/a_{\perp})$,
with $C=\left|\zeta(1/2)\right|/\sqrt{2}$~\cite{olshanii_atomic_1998,astrakharchik_quasi-one-dimensional_2004_short}.
As we see, the 1D interaction strength exhibits a ``geometric''
resonance for a certain 3D scattering length, that can be employed
to tune it by varying magnetic field. Here $a_{\perp}=(\hbar/m\omega_{0})^{1/2}$
is the harmonic oscillator length in the transverse direction for
the quasi-1D tubes with $\omega_{0}=(8\pi^{2}V_{0}/m\lambda^{2})^{1/2}$
. The number of particles in each tube is determined by the global
trap imposed on the 3D cloud. To justify the above continuum description
one should demand that all relevant energy scales of the system (collectively
denoted by $E$), which can be tuned to a great extent, obey the inequality
$\hbar\omega_{z}\ll E\ll\hbar\omega_{0}$.

Now, we begin by discussing the homogeneous situation \emph{i.e.},
$V_{\text{trap}}=0$. The model given by Eq.~(\ref{eq:Hamiltonian1D})
(and its lattice variant, known as the Hubbard model) has been exactly
solved by Bethe Ansatz~\cite{yang_exact_1967_combined}
and it is known that for repulsive interactions the low-energy and
long-wavelength physics is described by the SC Luttinger liquid, with
two copies of non-interacting charge and spin bosons, and is governed
by the Hamiltonian~\cite{gogolin_bosonization_2004}:
\begin{equation}
H_{\text{SC}}=\sum_{\alpha=c,s}\frac{v_{\alpha}}{2}\int dz\left[K_{\alpha}\Pi_{\alpha}^{2}+\frac{1}{K_{\alpha}}(\partial_{z}\varphi_{\alpha})^{2}\right],
\label{eq:HamiltonianLL}
\end{equation}
which describes the fluctuations of the charge and spin densities
above the ground state values. The bosonic field operators obey canonical
commutation relations $[\varphi_{\alpha}(z),\Pi_{\alpha'}(z')]=i\delta_{\alpha\alpha'}\delta(z-z')$
and $[\varphi_{\alpha}(z),\varphi_{\alpha'}(z')]=[\Pi_{\alpha}(z),\Pi_{\alpha'}(z')]=0$,
where $\Pi_{\alpha}=\frac{1}{K_{\alpha}v_{\alpha}}\partial_{t}\varphi_{\alpha}$
is the momentum conjugate to the bosonic field $\varphi_{\alpha}$
and $v_{c(s)}$ is charge (spin) velocity. For spin-rotation invariant
systems, the SC Luttinger liquid interaction parameter in the spin
sector is $K_{s}=1$~\cite{gogolin_bosonization_2004}. The spin
and charge velocities and the interaction parameter in the charge
sector can be determined from the exact solution for any interaction
strength and density~\cite{schulz_mesoscopic_1995}. In the weak
coupling regime, they can also be obtained by bosonization~\cite{gogolin_bosonization_2004}.
In contrast, the SI Luttinger liquid is described only by the charge
bosonic field~\cite{fiete_greens_2004,kakashvili_boundary_2007}.
The Hamiltonian in this regime, $H_{\text{SI}}$, can be easily obtained
from Eq.~(\ref{eq:HamiltonianLL}) by dropping the spin part and
making the substitutions: $K_{c}=1/2$ and $v_{c}=2v_{F}$ (which
is exact for local interactions). We assumed that the temperature
is very small compared to Fermi energy and considered the limit $J\rightarrow0$
first and then $T\rightarrow0$ (opposite order of limits than in
the SC Luttinger liquid). The behavior is equivalent to that of non-interacting
spinless fermions in agreement with the exact results~\cite{yang_exact_1967_combined}.

We shall now include the effects of the trap on a quasi-1D tube via
local density (Thomas-Fermi) approximation. Thus, the average density
of the system as a function of coordinate can be found as a solution
of the equation
\begin{equation}
\frac{dE[\left\langle \rho\right\rangle ]}{d\left\langle \rho\right\rangle }=\mu-V_{\text{trap}}(z),
\label{eq:LDA_LL}
\end{equation}
where $E[\left\langle \rho\right\rangle ]$ is the ground-state energy
of the uniform system per unit length and $\mu$ is the chemical potential
fixed by normalization, $\int\left\langle \rho(z)\right\rangle dz=N$.
In the SC Luttinger liquid regime, for weak interactions (near the
center of the trap)~\cite{recati_spin-charge_2003_combined_short}
\begin{equation}
E_{\text{SC}}[\left\langle \rho\right\rangle ]=\frac{\hbar^{2}\pi^{2}\left\langle \rho\right\rangle ^{3}}{24m}+\frac{U_{1D}\left\langle \rho\right\rangle ^{2}}{4},
\label{eq:E_LL}
\end{equation}
then from Eq.~(\ref{eq:LDA_LL}) the density reads
\begin{equation}
\left\langle \rho\right\rangle _{\text{SC}}(z)=\left\langle \rho_{0}\right\rangle \left(\sqrt{1+g^{2}-\frac{z^{2}}{R^{2}}}-g\right),\quad\left|z\right|\le R,
\label{eq:Density_LL}
\end{equation}
where $R=(2\mu/m\omega_{z}^{2})^{1/2}$, $\left\langle \rho_{0}\right\rangle =(8m\mu/\hbar^{2}\pi^{2})^{1/2}$
is the density of the non-interacting uniform system and $g=U_{1D}/\hbar\pi v_{F0}$,
with $v_{F0}=\hbar\pi\left\langle \rho_{0}\right\rangle /2m$. Due
to the coordinate dependence of the density of particles, the spin
and charge velocities, as well as the Luttinger liquid interaction
parameters become coordinate dependent~\cite{recati_spin-charge_2003_combined_short}.
From bosonization results we can write that $v_{s}(z)=v_{F}(z)-U_{1D}/2\pi\hbar$,
$v_{c}(z)=v_{F}(z)/K_{c}(z)$ and $K_{c}(z)=(1+U_{1D}/\hbar\pi v_{F}(z))^{-1/2}$.
These results are valid for weak coupling only, which is apparent
since the expressions break down near the trap edges where the interactions
become stronger. Since the trap potential does not break spin rotation
invariance, $K_{s}(z)=1$. For strong interactions, the SC Luttinger
liquid regime can only be realized near the trap center. Thus, we
will approximate $K_{c}(z)$ to its value at $z=0$, $K_{c}(0)=K_{c,0}$;
cf. Ref.~\cite{xia_effective_2005}.

We can do similar calculations for the spin-incoherent regime with
energy~\cite{recati_spin-charge_2003_combined_short}
\begin{equation}
E_{\text{SI}}[\left\langle \rho\right\rangle ]=\frac{\hbar^{2}\pi^{2}\left\langle \rho\right\rangle ^{3}}{6m},
\label{eq:E_SILL}
\end{equation}
since in this limit atoms behave as spinless particles. Using Eq.~(\ref{eq:LDA_LL})
again, the density reads this time
\begin{equation}
\left\langle \rho\right\rangle _{\text{SI}}(z)=\left\langle \rho_{0}\right\rangle \sqrt{1-\frac{z^{2}}{R^{2}}},\qquad\left|z\right|\le R,
\label{eq:Density_SILL}
\end{equation}
where $\left\langle \rho_{0}\right\rangle =(2m\mu/\hbar^{2}\pi^{2})^{1/2}$
is the density of the uniform non-interacting spinless system. For
the SI Luttinger liquid parameters we have: $K_{c}(z)=1/2$ and $v_{c}(z)=2v_{F}(z)$.
Taking the above into account the Hamiltonians for the SC Luttinger
liquid can be written as
\begin{equation}
H_{\text{SC}}\!=\!\!\!\sum_{\alpha=c,s}\!\frac{v_{\alpha,0}}{2}\!\!\!\!\!\!\int\limits _{-\tilde{z}_{\alpha}(R)}^{\tilde{z}_{\alpha}(R)}\!\!\!\!\!\! d\tilde{z}_{\alpha}\!\left[K_{\alpha,0}\tilde{\Pi}_{\alpha}^{2}+\frac{1}{K_{\alpha,0}}(\partial_{\tilde{z}_{\alpha}}\tilde{\varphi}_{\alpha})^{2}\right],
\label{eq:Hamiltonian_LDA_LL}
\end{equation}
where $d\tilde{z}_{\alpha}=dz/\tilde{v}_{\alpha}(z)$, $v_{\alpha}(z)=v_{\alpha}(0)\tilde{v}_{\alpha}(z)=v_{\alpha,0}\tilde{v}_{\alpha}(z)$,
$\tilde{\varphi}_{\alpha}(\tilde{z}_{\alpha})=\varphi_{\alpha}(z(\tilde{z}_{\alpha}))$
and $\tilde{\Pi}_{\alpha}(\tilde{z}_{\alpha})=\tilde{v}_{\alpha}(z(\tilde{z}_{\alpha}))\Pi_{\alpha}(z(\tilde{z}_{\alpha}))$.
One can easily check that the commutation relations of the new bose
fields remain canonical. $H_{\text{SI}}$ can be recovered from Eq.~(\ref{eq:Hamiltonian_LDA_LL})
proceeding as above. As we see, in the new coordinates the conformal
symmetry is restored. We can therefore use this powerful property
to calculate physical observables. Since the velocities vanish at
the trap edges, no particle and spin current can flow out of the system
and thus open boundary conditions (OBC) are effectively imposed on
the system.

Now we turn to calculate the matrix correlator of density fluctuations
which we shall show to differ qualitatively for different regimes
of the Luttinger liquid. The choice of this observable is motivated
by the versatility of trapped atomic systems, which give us the unique
opportunity to independently address spin components and study correlators
that are typically challenging to measure in solid state configurations~\cite{altman_probing_2004,folling_spatial_2005_short}.
The matrix correlation function that we are interested in is given
by
\begin{equation}
G_{\sigma\sigma'}(z,z')=\left\langle \delta\rho_{\sigma}(z)\delta\rho_{\sigma'}(z')\right\rangle ,
\label{eq:Def_Correlator}
\end{equation}
where $\delta\rho_{\sigma}(z)=\rho_{\sigma}(z)-\left\langle \rho_{\sigma}(z)\right\rangle =\delta\rho_{c}(z)/2+\sigma\delta\rho_{s}(z)$
is the fluctuation of the density of $\sigma$ species. Here we consider
only the smooth (non-oscillatory) part of the correlation function
and rewrite the correlator in spin and charge (\emph{i.e.}, total
particle) densities
\[
G_{\sigma\sigma'}(z,z')=\frac{1}{4}\left\langle \delta\rho_{c}(z)\delta\rho_{c}(z')\right\rangle +\sigma\sigma'\left\langle \delta\rho_{s}(z)\delta\rho_{s}(z')\right\rangle ,
\]
where we have used the property of spin-charge separation to drop
the correlations between the spin and charge densities. Using $\delta\rho_{c}(z)=\sqrt{2/\pi}\partial_{z}\varphi_{c}(z)$
and $\delta\rho_{s}(z)=\sqrt{1/2\pi}\partial_{z}\varphi_{s}(z)$,
the correlators are straightforwardly calculated. This is achieved
by using the conformal transformation $w'=(L/\pi)\ln w-iL/2$ ($L=2\tilde{z}_{\alpha}(R)$),
which maps the upper half complex plane ($w=v\tau+iz;\,0\le z$) (in
the imaginary time formulation), with an OBC imposed at $z=0$, to
a strip of width $L$ ($w'=v\tau'+iz';\,-L/2\le z'\le L/2$)~\cite{difrancesco_conformal_1997}.
Introducing the functions $\Delta_{\alpha}^{\pm}=\frac{\pi(\tilde{z}_{\alpha}(z)\pm\tilde{z}_{\alpha}(z'))}{4\tilde{z}_{\alpha}(R)}$
and $\tilde{R}_{\alpha}^{2}=64\tilde{z}_{\alpha}^{2}(R)\tilde{v}_{\alpha}(z)\tilde{v}_{\alpha}(z')$,
the correlation functions for the SC Luttinger liquid are given by
\begin{eqnarray}
G_{\sigma\sigma'}^{\text{SC}}(z,z') & = & G_{c}(z,z')+\sigma\sigma'G_{s}(z,z'),
\label{eq:Correlator_LL}
\end{eqnarray}
where
\begin{equation}
G_{\alpha}(z,z')=-\frac{K_{\alpha,0}}{\tilde{R}_{\alpha}^{2}}\left(\frac{1}{\sin^{2}\Delta_{\alpha}^{-}}+\frac{1}{\cos^{2}\Delta_{\alpha}^{+}}\right).
\label{eq:Def_correlator_LL}
\end{equation}

For the SI Luttinger liquid the average in the spin sector is zero
for large relative coordinates ($\left|z-z'\right|\gg a_{0}$, where
$a_{0}$ is the average distance between atoms), since the spin excitations
are exponentially damped~\cite{fiete_greens_2004} and the correlator
reads
\begin{equation}
G_{\sigma\sigma'}^{\text{SI}}(z,z')=-\frac{1}{2\tilde{R}_{c}^{2}}\left(\frac{1}{\sin^{2}\Delta_{c}^{-}}+\frac{1}{\cos^{2}\Delta_{c}^{+}}\right).
\label{eq:Correlator_SILL}
\end{equation}

Let us discuss these functions in more detail. Remarkably, the off-diagonal
correlators ($\sigma\ne\sigma'$) show a qualitative difference between
the SC Luttinger liquid and its spin-incoherent counterpart; namely,
$G_{\sigma\bar{\sigma}}^{\text{SC}}(z,z')>0$, while $G_{\sigma\bar{\sigma}}^{\text{SI}}(z,z')<0$.
(Notice that for a noninteracting system $G_{\sigma\bar{\sigma}}^{\text{NI}}(z,z')=0$.)

\begin{figure}
\includegraphics[width=3.2in,keepaspectratio]{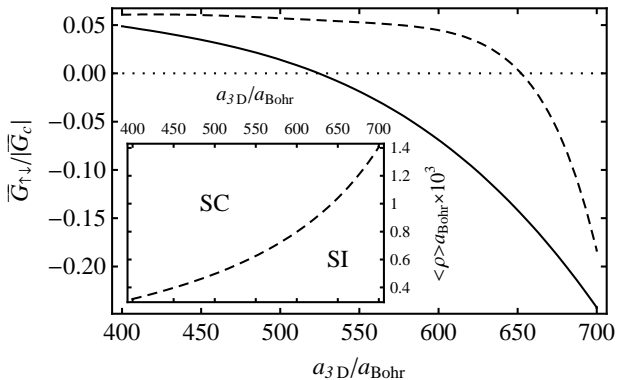} 
\caption{Qualitative behavior of $\bar{G}{}_{\uparrow\downarrow}/\left|\bar{G}_{c}\right|$
as a function of the 3D scattering length, $a_{\text{3D}}$, in a
harmonic trap for $^{6}\text{Li}$ atoms ($N=10^{5}$,$\omega_{\perp}=2160\pi\text{Hz}$,
$\omega_{0}=14\pi\times10^{5}\text{Hz}$ and $\omega_{z}=470\pi\text{Hz}$)
for the central tube (\emph{dashed}) and the array of tubes (\emph{solid}).
As the interaction strength is increased, the response changes to
negative, indicating the crossover from the spin-coherent Luttinger
liquid regime to its spin-incoherent counterpart. \emph{Inset}: ``Phase
diagram'' of a 1D ultra-cold atomic Fermi gas, the dashed curve is
a crossover line between the two regimes. }
\label{Correlator} 
\end{figure}

Having obtained the correlation function for a single tube we are
in a position to describe the response from the array of tubes, which
is more faithful to the experimental situation. Since we assume that
there is no particle transfer between the tubes, we can conclude that
they are uncorrelated. Thus, the many-tubes correlation function can
simply be written as
\begin{equation}
G_{\sigma\bar{\sigma}}^{ij}(z,z')=\left\langle \delta\rho_{\sigma}^{i}(z)\delta\rho_{\bar{\sigma}}^{j}(z')\right\rangle =\delta_{ij}G_{\sigma\bar{\sigma}}^{ii}(z,z'),
\label{eq:Def_Correlator_Multi}
\end{equation}
where the superscripts $i$ and $j$ label different tubes. The responses
from different tubes simply add up. In the \emph{in situ} experiment,
when the measuring laser beam probes the whole array of tubes, one
is able to measure the full integrated response, which is given by
$\bar{G}_{\sigma\bar{\sigma}}=\sum_{i}\int_{-R}^{R}G_{\sigma\bar{\sigma}}^{ii}(z,z')dzdz'$;
this averaging provides an observable with a good signal-to-noise
ratio. In Fig.~\ref{Correlator}, we show the qualitative behavior
of $\bar{G}_{\uparrow\downarrow}/\left|\bar{G}_{c}\right|$ for an
array of tubes confined in a global harmonic trap. We have used realistic
experimental parameters for $^{6}\text{Li}$ atoms to see that it
is possible to identify the response for the two regimes without pushing
the values to extreme limits. We see that the SI Luttinger liquid
regime is realized (as expected) for strong interactions and thus
yields the negative response, while the positive signal for weak interactions
is due to the SC Luttinger liquid correlations.

In conclusion, we have proposed an experiment to measure signatures
of strong interactions in 1D ultra-cold atomic systems. We have shown
that correlations of density fluctuations (an observable easily accessible
in atomic physics experiments compared to their condensed matter counterparts)
qualitatively distinguish between different strongly correlated regimes
in these systems and therefore provide an ideal probe for detecting
these regimes in cold-atom experiments. While the off-diagonal correlators
are positive for the spin-coherent Luttinger liquid, they are negative
for its spin-incoherent counterpart.
This result is robust, does not depend on the details
of the trap and should be easy to identify in the experiment.
This is also a novel proposal to measure properties
of the spin-incoherent regime in a non-condensed
matter system, which opens up new avenues for
the study of Luttinger liquids.
In particular, it would be interesting to extend
these studies to the spin-imbalanced case.

\begin{acknowledgments}
We thank the members of R. Hulet's group and especially G. Partridge
for illuminating discussions about the experimental setup. We would
like to acknowledge financial support from the W. M. Keck program
in Quantum Materials at Rice University, DARPA/ARO (W911NF-07-1-0464),
NSF and the Welch Foundation (C-1669). 
\end{acknowledgments}
\bibliographystyle{apsrev}
\bibliography{spin_inc}

\end{document}